\documentclass[a4paper,twocolumn,showpacs,superscriptaddress,prl]{revtex4}

\usepackage{graphicx}
\usepackage[english]{babel}
\usepackage{xspace}

\begin{document}

\title{Strain Relief in Cu-Pd Heteroepitaxy}

\author{Yafeng Lu}
\address{Max-Planck-Institut f\"{u}r Mikrostrukturphysik, Weinberg~2, D-06120 Halle, Germany}

\address{Northwest Institute for Nonferrous Metal Research, P. O. Box 51, Xian, Shaanxi 710016, P. R. China}

\author{M.~Przybylski}
\address{Max-Planck-Institut f\"{u}r Mikrostrukturphysik, Weinberg~2, D-06120 Halle, Germany}

\author{O.~Trushin}
\address{Institute of Microelectronics and Informatics of
Russian Academy of Sciences, Universitetskaya 21, Yaroslavl 150007, Russia}

\author{W.~H.~Wang}
\address{Max-Planck-Institut f\"{u}r Mikrostrukturphysik, Weinberg~2, D-06120 Halle, Germany}

\author{J.~Barthel}
\address{Max-Planck-Institut f\"{u}r Mikrostrukturphysik, Weinberg~2, D-06120 Halle, Germany}

\author{E.~Granato}
\address{Laborat$\acute{o}$rio Associado de Sensores
e Materiais, Instituto Nacional de Pesquisas Espaciais, 12245-970
S$\tilde{a}$o Jos$\acute{e}$ dos Campos, SP Brasil}

\author{S.~C.~Ying}
\address{Department of Physics,
Brown University, Providence, Rhode Island 02912, USA}

\author{T.~Ala-Nissila}
\address{Department of Physics,
Brown University, Providence, Rhode Island 02912, USA}

\address{Laboratory of Physics, Helsinki University of
Technology, FIN-02015 HUT, Espoo, Finland}


\date{February 4, 2005}

\begin{abstract}

We present experimental and theoretical studies of Pd/Cu(100) and
Cu/Pd(100) heterostructures in order to explore their structure
and misfit strain relaxation. Ultrathin Pd and Cu films are grown
by pulsed laser deposition at room temperature. For Pd/Cu,
compressive strain is released by networks of misfit dislocations
running in the [100] and [010] directions, which appear after a
few monolayers already. In striking contrast, for Cu/Pd the
tensile overlayer remains coherent up to about 9 ML, after which
multilayer growth occurs. The strong asymmetry between tensile and
compressive cases is in contradiction with continuum elasticity
theory, and is also evident in the structural parameters of the
strained films. Molecular Dynamics calculations based on classical
many-body potentials confirm the pronounced tensile-compressive
asymmetry and are in good agreement with the experimental data.

\end{abstract}












\maketitle


Heteroepitaxy with different thicknesses ranging from monoatomic
layers up to micrometers produces various structures, and is one
of most important routes to artificially obtain functional
materials and devices. Stress and its relaxation due to an
interface lattice mismatch are an essential problem in the
heterostructures. Classical continuum theory defines the concept
of an equilibrium critical thickness $h_{c}$, above which the
lattice stress relaxes with the introduction of a misfit
dislocation (MD)~\cite{CCM}. The critical thickness is determined
by energy balance between strain energy buildup and strain relief
due to dislocation nucleation in the mismatched structure. The
value of $h_{c}$ predicted by the continuum theory, however, often
does not agree with experiments. Moreover, the critical thickness
in the continuum elasticity theory is independent on the strain
type, be it tensile or compressive. Atomic details of the
interface structure such as surface steps and surface roughness,
are usually ignored. Actual stress relaxation of coherent strained
films is kinetically limited, typically with high barriers.
Recently, atomistic studies on model systems have been done to
determine the transition paths and energy barriers from coherent
to incoherent states in real epitaxial
films~\cite{Dong:98,Trushin:02,Trushin:03}. One of the main
findings has been a striking asymmetry between the tensile and
compressive cases, due to the anharmonicity and asymmetry of the
atomic interactions~\cite{Trushin:02,Trushin:03}. Further, the
concept of a size-dependent mesoscopic mismatch has been proposed
to explain strain relaxations in the early stage of homo and
heteroepitaxial metal growth on the atomic
scales~\cite{Kirschner}.

In this Letter, we present results from a combined experimental
and theoretical study on a model system of Cu-Pd heteroepitaxy
grown by pulsed laser deposition (PLD). The bulk lattice
parameters of Cu and Pd are $a_{\tiny{\textrm{Cu}}}=3.61$\AA ~and
$a_{\tiny{\textrm{Pd}}}=3.89$\AA, respectively. The lattice misfit
induces a large compressive ($m=-7.2\%$) strain in the Pd
overlayer on Cu(100) and a tensile ($m=7.8\%$) strain for the Cu
overlayer on Pd(100). The Pd films on Cu(100) and the Cu films on
Pd(100) were grown by pulsed laser deposition (PLD) in a
multi-chamber ultrahigh vacuum (UHV) system with a base pressure
$p < 5\times\,10^{-11}$ mbar, and $p < 2\times\,10^{-10}$
mbar during deposition. Prior to deposition the copper and
palladium substrates were cleaned by cycles of Ar$^{+}$ sputtering
followed by annealing at $873$ K for Cu substrate or $950$ K for
Pd substrate until clean Auger electron spectroscopy (AES)
spectra, sharp low-energy electron diffraction (LEED) spots, and
atomically smooth terraces under scanning tunneling microscopy
(STM) were observed. The substrate temperature was kept at room
temperature during deposition. The substrate was placed about
$100\sim 130$ mm away from targets. A KrF excimer laser beam
(wavelength $248$ nm, pulse duration $34$ ns, typical pulse energy
$300-350$ mJ and pulse repetition rate of $3-5$ Hz) was focused
onto targets. During deposition the growth process was monitored
by a reflection high energy electron diffraction (RHEED). All STM
measurements were performed in the constant current mode at a
$0.2-0.5$ V positive tip bias and a $0.1-0.5$ nA tunneling
current. Compared to thermal deposition (TD), an extremely high
instantaneous flux of atoms in PLD favors much larger nucleation
density~\cite{Jenniches1:99,Ohresser:99}. This helps to suppress
the formation of Cu(001)$c(2\times 2)$-Pd ordered alloy in the
initial growth stage for Pd/Cu(100)~\cite{Murray:95} and the phase
transition at 1 ML from fcc to bct structure for
Cu/Pd(100)~\cite{Hahn:95}.
Even though the present STM setup does not provide a direct
distinction of the Pd and Cu atoms,  neither pits in the substrate
surface nor eroded substrate step edges show any evidence of Cu-Pd
interlayer mixing. Moreover, the surface area covered by the
islands for submonolayer coverages, as determined from STM
images, are always consistent with the coverage from RHEED.
The RHEED data confirm the layer-by-layer growth mode for both systems.
It persists up to at least $5-6$ ML in Cu-Pd heteroepitaxy.

\begin{figure}[t]
\includegraphics[bb= 7cm  17cm  15cm   24cm, width= 8cm]{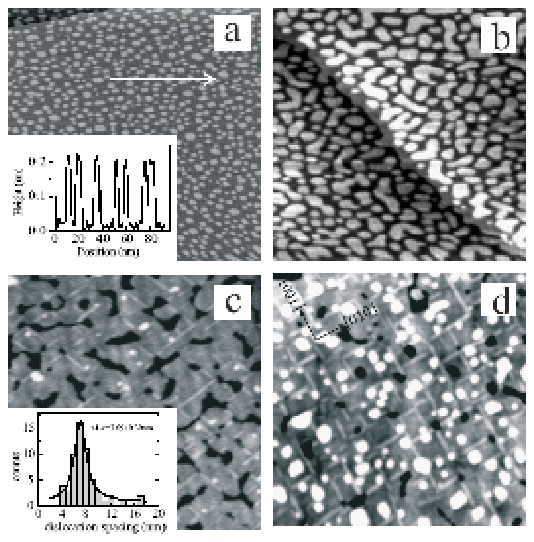}
 \caption{\small STM images for (a) $0.5$ ML Pd ($200\times 200$ nm$^{2}$);
 (b) $1.6$ ML Pd ($200\times 200$nm$^{2}$); (c) $4.8$ ML Pd ($100\times 100$
 nm$^{2}$); (d) $6.0$ ML Pd ($100\times 100$
 nm$^{2}$ deposited at $300$ K.}
 \label{STMPdCu}
 \end{figure}

Figures \ref{STMPdCu}(a)-(d) display STM images taken at room
temperature for Pd films on Cu(100). For submonolayer Pd at $0.5$
ML, many islands with a monolayer height of about $1.80$ {\AA}
were observed as shown in the inset of Fig.~\ref{STMPdCu}(a). With
increasing Pd thickness such islands are always formed on a
completely filled underlayer up to a total thickness of about
$3-4$ ML, indicating ideal 2D growth mode. Above $4$ ML a striking
feature is the appearance of misfit dislocations as marked by many
protruding stripes with an average height of $\Delta h = (0.57\pm
0.10)$\AA~ (see Fig.~\ref{STMPdCu}(c) and (d)), by which the
compressive stress in Pd film relaxes efficiently. They align
nearly regularly along the [100] and [010] directions. The average
spacing of MD lines is $(7.09 \pm 0.30)$~nm for $4.8$ ML. The
layer-by-layer growth persists up to $6$ ML, after which
multilayer growth mode occurs.

 \begin{figure}[t]
\includegraphics[bb= 3cm  17cm  15cm   23cm, width= 8cm]{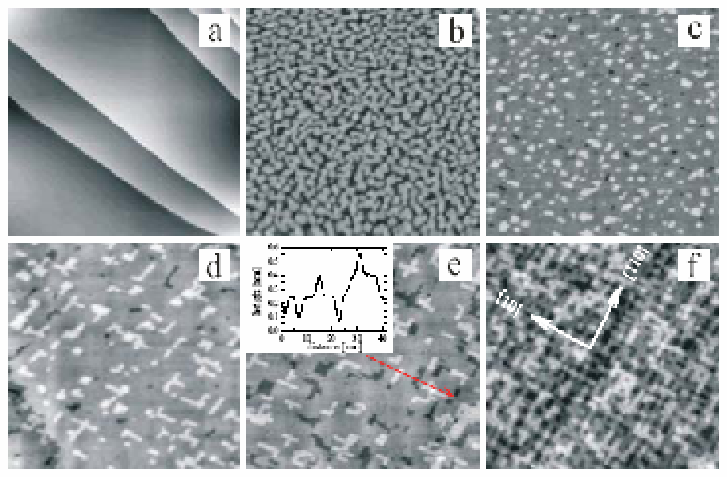}
 \caption{\small STM images for (a) clean Pd(100)
 substrate ($500\times 500$ nm$^{2}$); (b)
 $0.85$ ML Cu ($100\times 100$ nm$^{2}$); (c) $2.1$ ML Cu ($100\times 100$
 nm$^{2}$); (d) $3.05$ ML Cu ($100\times 100$
 nm$^{2}$); (e) $6.0$ ML Cu ($100\times 100$
 nm$^{2}$); (f) $9.0$ ML Cu ($100\times 100$
 nm$^{2}$) deposited at $300$ K.}
 \label{STMCuPd}
 \end{figure}

The corresponding layer-by-layer growth mode for PLD Cu films on
Pd(100) has also been confirmed by STM images as indicated in Fig.
\ref{STMCuPd}. At $0.85$ ML, second layer nucleation is not found,
while in the $2.1$ ML thick film more than $95$\% of the second
layer is accomplished (Figs.~\ref{STMCuPd}(b) and (c)). At $2.1$
ML, few of the islands in the third layer have a preferential
orientation, whereas at $3.05$ ML most of the fourth layer islands
are rectangular along the $\langle 110 \rangle$ directions. With
increasing Cu thickness to $6.0$ ML, $90$\% of the sixth layer is
completed, where nearly all the pits in the sixth layer and the
seventh layer islands are rectangular. Above $6$ ML the morphology
of Cu films is changed because of the multilayer growth mechanism.
A typical image for $9.0$ ML is given in Fig.~\ref{STMCuPd}(f).
Four layers appear simultaneously on the surface. Moreover, only
$70$\% of the ninth layer is completed while the remaining $30$\%
appears as the tenth and eleventh layers. It should be pointed out
that the height or depth of all (ir)regular islands and pits is
about $1.65$ {\AA} until $9.0$ ML as shown in the inset of
Fig.~\ref{STMCuPd}(d). This vertical magnitude corresponds to
the interlayer spacing of highly strained fcc Cu~\cite{Li:91}. In
PLD Cu films on Pd(100) we have not found any trace of dislocation
nucleation up to $9$ ML. Above $9$ ML, the multilayer island
structure with the $\langle 110 \rangle$ orientation is still
visible and the film surface becomes more and more rough.


\begin{figure}[t]
  \includegraphics[bb= 3cm  8.5cm  18cm   20cm, width= 8cm]{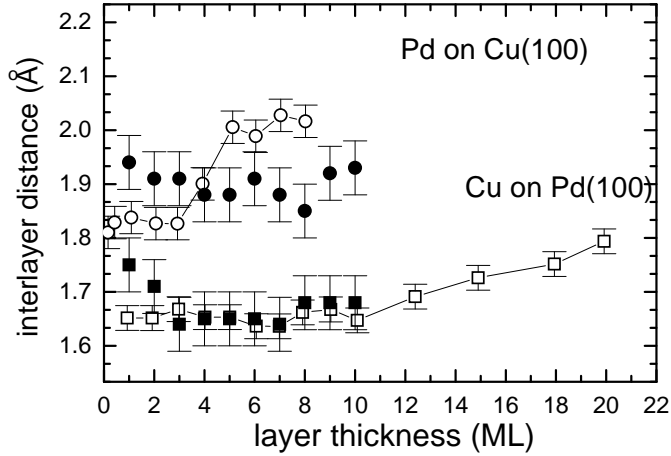}


 \caption{\small Perpendicular interlayer distance calculated
 based on a kinetic model from LEED intensity of the (00) diffraction beam
{\it vs.} energy curves. The change of the interlayer distance
occurs at about $3-4$ ML for Pd/Cu(100) films and at about $10$ ML
for Cu/Pd(100) films. Assuming a pseudomorphic growth up to $3$ ML
for Pd on Cu(100) and $10$ ML for Cu on Pd(100), PLD films can be
identified as fcc Pd and fct Cu structures. The results from
experimental and theoretical work at $300$ K are indicated by open
and solid symbols, respectively. The lines are guides to eye.}
 \label{leed}
 \end{figure}

IV-LEED measurements of the specularly scattered electron beam
were done to obtain structural information on the films. The
average interlayer spacing of PLD films was calculated from LEED
intensity of the (00) diffraction beam {\it vs.} energy curves
based on a kinetic model~\cite{Zharnikov:96}. Fig.~\ref{leed}
gives the layer thickness dependence of the interlayer distance
for both systems. In the case of Pd/Cu(100) films, the interlayer
distance is about $1.82(5)$~\AA~ up to $3$ ML, very close to the
value of Cu substrate, and increases rapidly above $3$ ML to reach
the value of bulk Pd. For Cu on Pd(100) up to $10$ ML, the value
stays constant ($\approx 1.64$~\AA). This agrees well with the
measured height of monolayer islands by STM. Above $10$ ML the
interlayer spacing gradually increases close to the value of bulk
fcc Cu at a coverage of $20$ ML.

\begin{figure}[b]
 \includegraphics[bb= 3cm  8cm  18cm   20cm, width= 8cm]{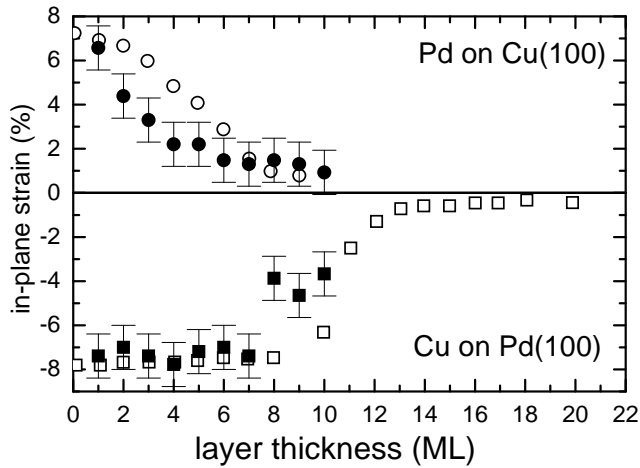}
 \caption{\small In-plane lattice strain derived from the separation of
 the first order rods in the RHEED pattern as a function
 of Pd or Cu coverages. Our results from experimental and
 theoretical work at $300$ K are indicated by open and
 solid symbols, respectively. See text for details.}
 \label{strain}
 \end{figure}

The actual in-plane lattice constant of deposited films was
obtained \textit{in situ} by measuring the spacing between the
reciprocal lattice rods in the RHEED image. The RHEED spacing of
substrates serves as a reference. The in-plane lattice strain is
defined as
$(a_{\tiny{\textrm{bulk}}}-a_{\tiny{\textrm{film}}})/a_{\tiny{\textrm{bulk}}}$(\%).
Fig.~\ref{strain} gives the measured data. For Pd/Cu(100), the
measured in-plane strain exhibits a gradual decrease with
increasing thickness of the Pd film.
The compressive strain is nearly fully relaxed at $9$ ML. In
contrast, the in-plane tensile strain shows a very different
thickness dependence in the case of PLD Cu films on Pd(100). Until
$9$ ML only $0.6$ \% reduction in the tensile strain is found.
Above $9$ ML a step-like decrease is visible and a residual strain
is observed until $20$ ML.
\begin{figure}[t]
 \includegraphics[bb= 3cm  20cm  15cm   24cm, width= 8cm]{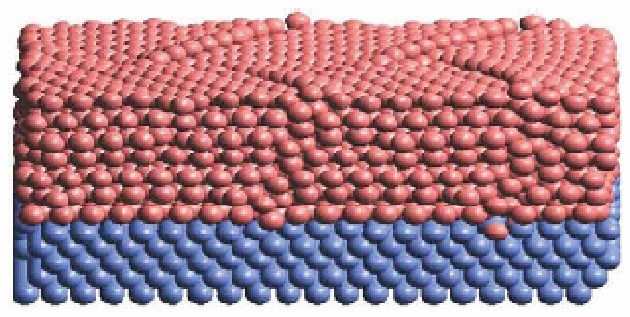}
 \caption{\small Typical configuration of stacking faults for $7$ ML
 Pd on Cu(100) based on
 molecular dynamics annealing. The lateral size is $20
\times 20$ atom$^{2}$. The misfit dislocations orientate along the
$\langle 100 \rangle$ direction of the substrate.}
 \label{defect}
 \end{figure}


The most striking discovery in our data is the strong asymmetry in
the strain relaxation behavior between the tensile and compressive
strained films. This result is in direct disagreement  with the
continuum elasticity theory, which predicts a symmetrical behavior
for tensile and compressive systems.

In any growth mode, there is always the question whether the final
configuration is dictated by kinetic considerations or it
corresponds to a true thermodynamic equilibrium state. Given the
lack of precise characterization of the PLD growth conditions, a
full theoretical modeling of the kinetic growth process under the
experimental conditions is unfeasible. Instead, we have chosen to
perform equilibrium molecular dynamics simulations of the
epitaxial layer after deposition. The object here is to study the
thermal excitation and nature of dislocations in the epitaxial
film and to see whether they can account for the observed stress
relaxation and tensile compressive asymmetry in Cu-Pd
heterostructures.
Our atomistic model contains five layers of substrate and varying
number of film layers. Two bottom layers of the substrate were
fixed. Periodic boundary conditions were applied in the plane
parallel to the surface. The lateral size of the systems studied
varied between $20\times20$ to $40\times40$. Interatomic forces
between particles in the system were computed using Embedded Atom
Model (EAM) potentials for Cu and Pd~\cite{eam}. The initial
configuration of the film has adatoms occupying fcc positions in
registry with substrate.

For Pd/Cu(100), after thermal equilibrium is reached at $300$ K,
the system relaxes and gains partial strain relief through
generation of misfit dislocations starting at $2$ ML and above. To
facilitate a direct comparison with the experimental data in Fig.
\ref{leed}, we computed from our simulations the average
interlayer distance for the top layers of the film (up to three to
take into account the surface sensitivity of LEED). We also
computed the average in-plane lattice constant for the top three
surface layers, and then evaluated the in-plane strain using the
bulk value of the adsorbate as a reference. The dependence of
these values on the film thickness is shown in Fig. \ref{strain}.
There is a good qualitative agreement between the theoretical
results with the experimental observation showing continuous
relief of in-plane strain through the misfit dislocations. The
exact nature of the misfit dislocations is rather complicated,
varying both as a function of film thickness and lateral size. In
Fig.~\ref{defect}, we show a typical configuration for a film of
$7$ ML thickness and lateral size $20 \times 20$ obtained by
molecular dynamics annealing through heating to $500$ K, followed
by cooling to $0$ K. This shows a Pd film with misfit dislocations
visibly aligned along the $\langle 100 \rangle$ direction of the
substrate, in agreement with the experimental observations.

Simulation studies were also performed for Cu/Pd(100). Here, in
agreement with the experimental observation, the theory shows a
striking difference from the Pd/Cu(100) system. The pseudomorphic
strained fct structure for the Cu film remains stable up to about
$9$ ML. Then strain is gradually released through localized
surface defect such as vacancies and adatoms, rather than the
misfit dislocation mechanism observed in Pd/Cu(100). The
theoretical results for the average interlayer distance of the
film and the in-plane strain for the top surface layers as a
function of the film thickness are shown in Figs. \ref{leed} and
\ref{strain}, respectively. The theory again is in good agreement
with the experimental data, indicating a stable highly strained
epitaxial state below $9$ ML.


In conclusion, we have studied the stress relaxation mechanism in
Cu-Pd heteroepitaxy both through PLD experiments and numerical
simulations. The experimental data show that for Pd/Cu(100), the
in-plane stress is relaxed almost immediately above $1$ ML,
whereas Cu/Pd(100) remains coherent up to about $9$ ML. While
these observations contradict predictions from the continuum
theory, they are confirmed by our atomistic simulations using EAM
potentials for the Pd and Cu interactions. The good agreement
between the experimental data and the equilibrium simulation study
leads us to conclude that the observed configurations of the film
at different thicknesses correspond to equilibrium state with
thermally generated misfit dislocations as the strain relief
mechanism. The microscopic nature of the misfit dislocation
observed here for the Pd/Cu(100) system is rather complicated and
does not fit into the simple model postulated
previously~\cite{muller}.

On a microscopic level, the tensile-compressive asymmetry
originates from the asymmetry of the interatomic interactions. The
general feature of interatomic potentials is a steeply rising
repulsive part at small separations as opposed to a very slowly
decreasing attractive part at large separations. This strong
asymmetric feature of the interaction is what leads to the
observed macroscopic tensile-compressive asymmetry. We have
performed additional studies using very different potentials (of
Lennard-Jones type) with the same general feature, and the results
indeed remain qualitatively unchanged. Interestingly, we find that
geometry and dimension play an important role in the
tensile-compressive strain relief asymmetry. For the 2D model,
misfit dislocations are favored for tensile rather than
compressive strained systems~\cite{Trushin:02,Trushin:03}.
Clearly, there is much more to learn in strain relief in
heteroepitaxy through further theoretical and experimental
investigations.

\bigskip

Y. Lu acknowledges support from the MPI in Halle during his stay
there. This work has been supported in part by Funda\c c\~ao de
Amparo \`a Pesquisa do Estado de S\~ao Paulo - FAPESP (grant no.
03/00541-0) (E.G.), the Academy of Finland through its Center of
Excellence program (T.A-N. and O.T.), and CRDF grant
RU-P1-2600-YA-04 (O.T.).


\end{document}